\documentclass[
apl,
reprint,
amsmath,amssymb,
superscriptaddress 
]{revtex4-1}
\usepackage{setspace}
\usepackage{graphicx}
\usepackage{float}
\usepackage{ragged2e}
\usepackage{xcolor}
\usepackage{caption}
\DeclareCaptionJustification{justified}{\justifying}
\usepackage[caption=false]{subfig}

\usepackage{bm} 
\usepackage{hyperref}

\setcitestyle{super}

\begin{document}

\title{{Quantum random number generation using spatial quantum noise of light}}


\author{Mohamed Armoon Shaliq}
\affiliation{Department of Physics, Indian Institute of Space Science and Technology, Thiruvananthapuram, Kerala 695547, India}
\author{Ashok Kumar}
\email{ashokkumar@iist.ac.in} 
\affiliation{Department of Physics, Indian Institute of Space Science and Technology, Thiruvananthapuram, Kerala 695547, India}

\begin{abstract}
Generating high-speed, verifiable random numbers is a fundamental requirement for cryptography, large-scale stochastic simulations, and secure quantum communication. Here, we present a robust quantum random number generator that utilizes the spatial distribution of quantum fluctuations captured by an electron-multiplying charge-coupled device operated in high-speed kinetic mode. Through rigorous detector calibration and shot-noise analysis, we characterize the spatial quantum noise obtained from the spatial intensity fluctuations of the coherent states of light. Such quantum noise serves as a high-entropy source, enabling an instantaneous random bit generation rate of 5.92 Gbps without algorithmic randomness extraction in the present configuration. The sustained output rate is, however, limited to 7.5 Mbps by the bandwidth of the serial electronic readout. The generated sequences successfully pass the NIST SP 800-22 and Marsaglia Diehard statistical test suites, confirming the high quality and unpredictability of the entropy source. 
\end{abstract}

\maketitle
High entropy random numbers are essential for secure cryptographic protocols, banking systems, and stochastic  simulations \cite{QM}. These random numbers can be generated using computer algorithms called pseudo-random number generators which are deterministic and vulnerable to seed cracking, posing significant security risks \cite{schneier1996applied}. To address these limitations, true random number generators harvest entropy from macroscopic physical processes like thermal noise in electronic circuits\cite{noise,petrie2000noise}, though they remain theoretically predictable under Newtonian physics. On the other hand, quantum random number generators (QRNGs) overcome this predictability by deriving true randomness directly from the intrinsic quantum fluctuations\cite{bera2017randomness,ma2016quantum}. While early QRNGs relied on radioactive decay\cite{manelis1961generating}, the requirement for a highly radioactive source and the low bit rate made these systems impractical for scalable and commercial random number generation\cite{schmidt1970quantum}. The modern commercial QRNGs extract entropy from light \cite{TU}, leveraging its speed and compatibility with existing fiber-optic infrastructure and quantum key distribution applications\cite{ani}.

While some protocols of optical QRNG are based on single photon detection event at a beam splitter \cite{jennewein2000fast} to generate randomness. The fundamental complexity of generating deterministic single photons \cite{SP,SI,QD} and the presence of inherent technical biases impose strict limitations on their generation rates and scalability. Consequently, practical implementations shifted to attenuated lasers producing weak coherent states\cite{collantes2017quantum,COH,ren2011quantum}, in which the photon statistics is governed by the Poissonian distribution. The inherent Poissonian fluctuations of these states provide a robust entropy source \cite{PP,WC,QNG}, overcoming the low count rates of true single-photon generation. Moreover, one can extract the random numbers from the continuous quadrature measurement of the coherent state\cite{gabriel2010entropy,grosshans2002continuous, gerry2022proposal}. The uncertainty in such field, manifesting as intrinsic shot noise, scales with the square root of the photon number and supports high generation bandwidths, enabling chip-level integration \cite{SIL,regazzoni2019high,sanguinetti2014quantum}. To date, most modern QRNGs rely on strictly temporal measurements, making them sensitive to timing jitter \cite{UF, ZU} and often requiring use of cryptographic randomness extractors that causes severe entropy loss and reduced generation speeds\cite{AL}.

Here, we present a extractor-free QRNG based on the extraction of spatial quantum noise from the coherent states, using an electron-multiplying charge-coupled device (EMCCD) camera. We utilize the spatial quantum optical intensity fluctuations across the two-dimensional transverse plane of the optical beam \cite{kumar2017observation}. The quantum fluctuations are digitized into a random bit stream and subjected to rigorous statistical evaluation without use of any cryptographic randomness extractors. To maximize the bit extraction rate, two parallel beams of equal optical power derived from a single coherent state are simultaneously detected by a single EMCCD. The quantum origin and inherent randomness are verified through spatial cross-correlation and shot noise analysis of the spatial fluctuations. 

In what follows, we begin with a brief discussion on coherent state statistics, followed by a detailed description of the experimental setup, noise analysis methodology and random bit extraction. We conclude our study with statistical validation of the random numbers by using NIST SP 800-22 test suite and Diehard statistical test.

\begin{figure*}[t]
    \centering
    \includegraphics[width=1.0\textwidth]{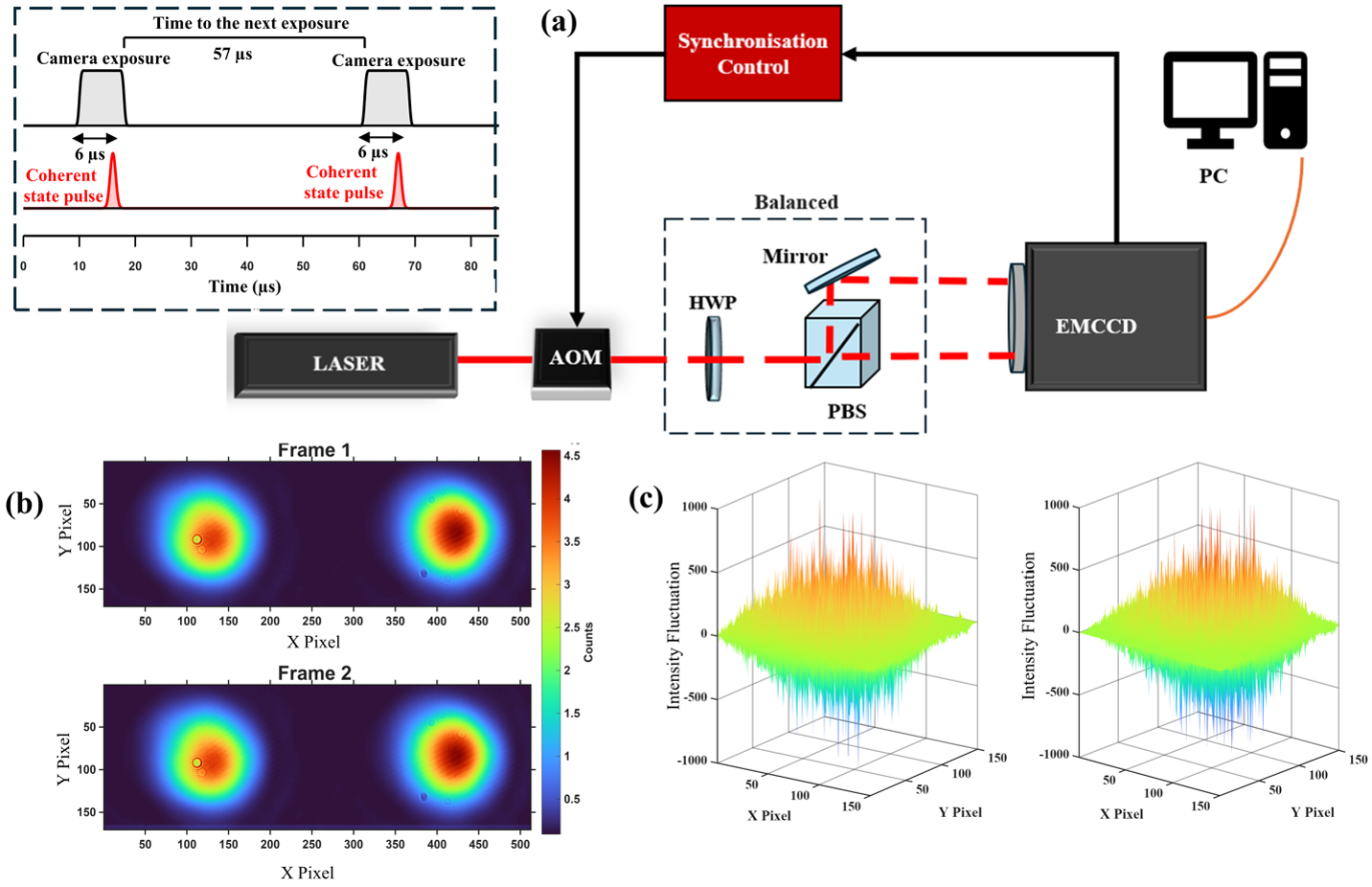}
    
    \caption{(a) Experimental setup for extracting spatial quantum noise from a coherent state (details are given in the main text). The inset illustrates the timing sequence, where the coherent state pulse is synchronized with the EMCCD exposure time of $10~\mu\text{s}$. (b) Representative EMCCD frames of the left and right coherent state pulses, and (c) spatial quantum fluctuation of respective left and right coherent states obtained via subtraction of two consecutive frames shown in (b).}
    \label{fig:exp_setup}
\end{figure*}

An ideal single-mode laser emits light in a coherent state $|\alpha\rangle$, representing the quantum state that most closely approximates a classical electromagnetic wave while possessing minimum uncertainty product. In the photon-number (Fock) basis, $|n\rangle$, the coherent state is expressed as,
\begin{equation}
    |\alpha\rangle = e^{-\frac{|\alpha|^2}{2}} \sum_{n=0}^{\infty} \frac{\alpha^n}{\sqrt{n!}} |n\rangle,
\end{equation}
where $n$ is the number of photons and $\alpha$ is a complex amplitude whose squared magnitude, $|\alpha|^2$, corresponds to the mean photon number $\langle n\rangle$. 
The probability $P(n)$ of detecting exactly $n$ photons in a coherent state is,
\begin{equation}
    P(n) = \frac{|\alpha|^{2n} e^{-|\alpha|^2}}{n!} = \frac{\langle n \rangle^n e^{-\langle n \rangle}}{n!},
\end{equation}
 which corresponds to a Poissonian distribution in photon number having variance exactly equal to the mean, i.e, $(\Delta n)^2 = \langle n \rangle$. This statistical fluctuation, commonly referred to as the shot noise, is an unpredictable property of the quantum state, serves as the core physical entropy source in the current investigation.
 
 The experimental configuration and the corresponding timing sequence of the coherent state pulses for spatial quantum noise extraction are illustrated in Fig.~\ref{fig:exp_setup}(a). A Titanium-Sapphire laser operating at 795 nm is pulsed with 1 $\mu$s width using an acousto-optic modulator (AOM). The generated pulse propagates through a half-wave plate (HWP) to precisely control the polarization state before being split by a polarizing beam splitter (PBS) into two equal power pulses.  This combination ensures equal optical power and identical mean photon flux in both coherent state pulses, providing the balanced operating condition required for the subsequent noise analysis. The parallel and balanced power pulses, labeled as left and right coherent pulses, are acquired by the EMCCD camera and recorded in a PC. As shown in the inset of Fig.~\ref{fig:exp_setup}(a), the $1~\mu\text{s}$ coherent state pulse is actively synchronized with the EMCCD camera’s $10~\mu\text{s}$ exposure signal. Specifically, the coherent state pulse is triggered $6~\mu\text{s}$ after the start of the exposure window to ensure it falls strictly within the acquisition period.
 We utilize an EMCCD camera in kinetic mode where we divide the total active sensor area, which consists of 512 × 512 pixels and an additional 512 × 512 pixels buffer region for storage, into six frames of $170 \times 512$ pixels each. During the acquisition sequence, only the topmost frame is exposed and the resulting charges are then shifted vertically at a rate of 300~ns/row which creates a $51~\mu\text{s}$ transfer window for 170 rows (a single frame). When combined with the exposure time, it results in a total temporal separation of 57~$\mu$s between the consecutive frames. In Fig.~\ref{fig:exp_setup}(b), we show the images acquired by two consecutive frames of the camera. This cycle continues until a batch of six frames is stored on-chip and read out collectively. Employing this kinetic acquisition mode minimizes readout dead time and maximizes the stability of the high-speed data capture. 
 
 In order to extract the spatial quantum noise from the acquired images, the intensity profiles from two consecutive frames are subtracted which effectively cancels out the static classical DC profile leaving behind the pure spatial quantum noise\cite{bg} as shown in Fig.~\ref{fig:exp_setup}(c). To systematically eliminate electronic noise, we acquire a corresponding background frame in the absence of the coherent pulse immediately following each signal capture and subtracting them from the corresponding acquired coherent state pulse images.
 To verify the Poissonian statistics of the two extracted spatial quantum fluctuations, we calculate a spatial cross-correlation coefficient between them as,
\begin{equation}
    C(i, j) = \frac{\sum_m \sum_n \delta I_L(m,n) \delta I_R(m-i, n-j)}{\sqrt{\left(\sum_m \sum_n \delta I_L^2(m,n)\right) \left(\sum_m \sum_n \delta I_R^2(m,n)\right)}},
    \label{eq:cross_corr}
\end{equation}
 where $\delta I_L(m,n)$ and $\delta I_R(m,n)$ represent the spatial quantum fluctuations for the left and right coherent state pulses, respectively, obtained by subtracting two consecutive frames. The summation extends over all pixels within the defined region of interest (ROI). 
 
\begin{figure}[t]
    \begin{minipage}{0.48\columnwidth}
        \raggedright \textbf{(a)} \\
        \vspace{0.2em}
        \includegraphics[width=\linewidth]{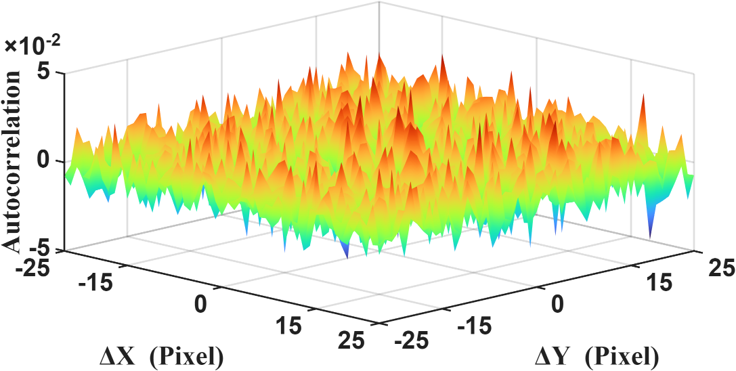}
    \end{minipage}
    \hfill 
    \begin{minipage}{0.48\columnwidth}
        \raggedright \textbf{(b)} \\
        \vspace{0.2em}
        \includegraphics[width=\linewidth]{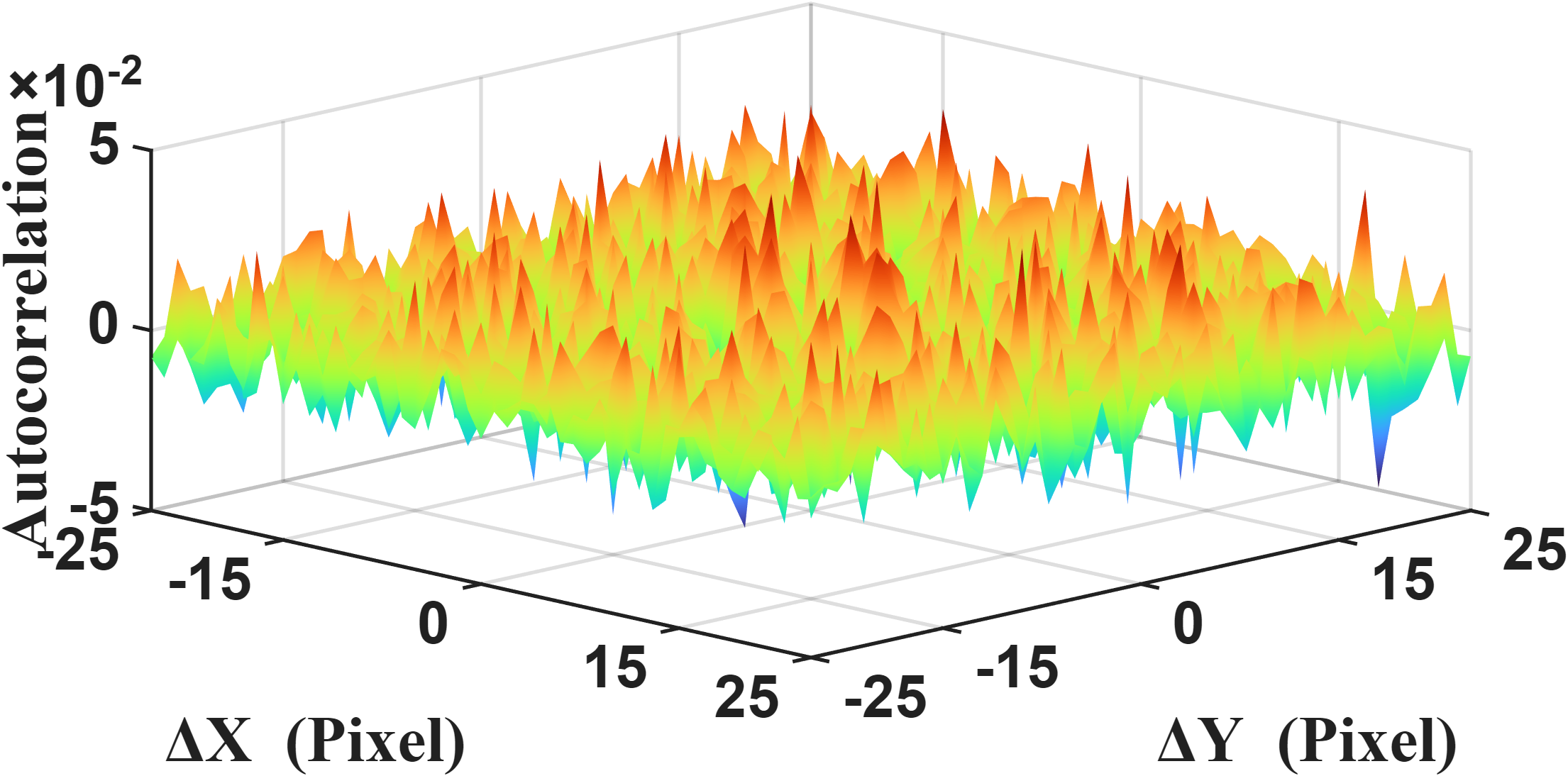}
    \end{minipage}

    \begin{minipage}{1.0\columnwidth}
        \raggedright \textbf{(c)} \\
        \vspace{0.2em}
        \includegraphics[width=\linewidth]{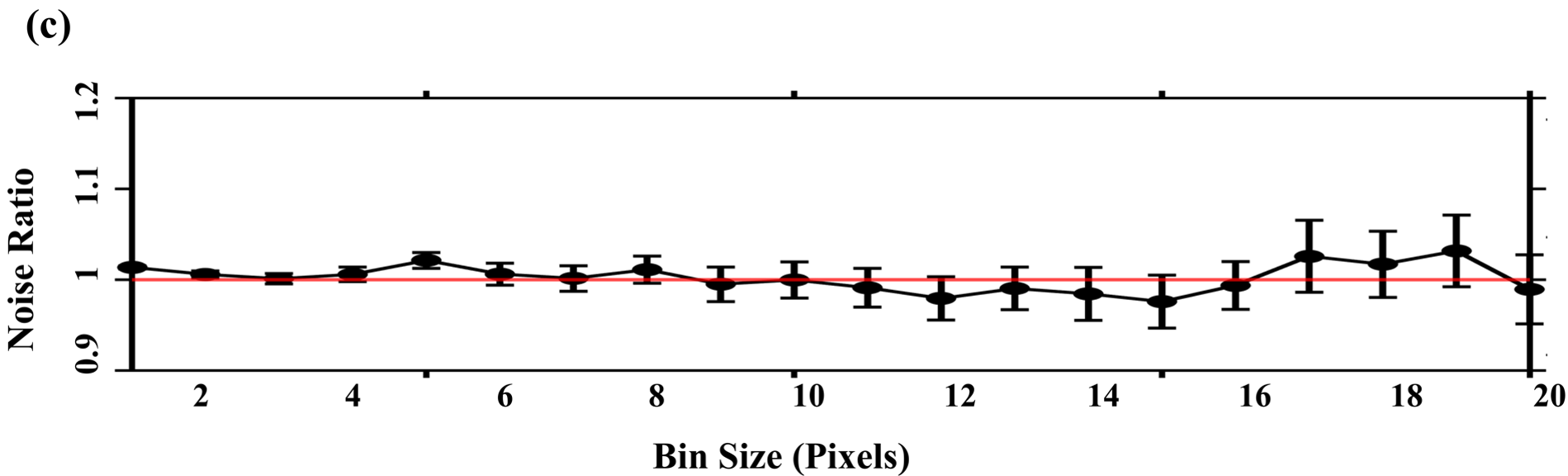}
    \end{minipage}    
    \caption{(a) Spatial cross-correlation map between the left and right coherent state intensity fluctuations. (b) Spatial auto-correlation of a single coherent state intensity fluctuation, the artificial self-correlation peak value is kept to zero. (c) Measured noise ratio (NR) for coherent state as a function of binning. Error bars represent the standard deviation of the mean for the NR over 100 acquired images.}
    \label{fig:purity_analysis}
\end{figure}

 We extract an ROI of $150 \times 150$ pixels centered on the intensity maximum of the left pulse's fluctuation image, alongside a smaller $100 \times 100$ pixel ROI for the right side pulse image. The extracted right pulse image ROI is then scanned across the left pulse image ROI to compute the spatial cross-correlation map. This process is computationally equivalent to a two-dimensional convolution where the kernel (the right pulse ROI) is not spatially inverted. The resulting map is plotted in Fig.~\ref{fig:purity_analysis}(a), absence of any correlation peak confirms that the two spatial quantum noise distributions are statistically independent and random.  Furthermore, to verify spatial independence in intensity fluctuation across a single pulse, we perform a spatial auto-correlation measurement using the same mathematical expression as the cross-correlation given in Eq.~\ref{eq:cross_corr}, except that the correlation is calculated between spatial fluctuations within a single beam rather than between the left and right beams. Thus, $\delta I_L(m,n)$ and $\delta I_R(m,n)$ in Eq.~\ref{eq:cross_corr}  are replaced by the fluctuations of the same beam. As shown in Fig.~\ref{fig:purity_analysis}(b), the correlation coefficients across 100 pixels of spatial scans remain randomly distributed near zero. This confirms that each pixel acts as a statistically independent entropy source, free from detector-induced artifacts or spatial cross-talk. While the autocorrelation analysis confirms the absence of significant spatial correlations within each beam, the cross-correlation analysis verifies the statistical independence of the two beams recorded on the separate halves of the EMCCD. Together, these results confirm the spatial randomness of the intensity fluctuations across all frames used for random bitstream extraction. Furthermore, low-frequency classical noise are effectively suppressed through frame subtraction, low-noise operation, and deep thermoelectric cooling, resulting in no measurable residual spatial autocorrelation. Therefore, within experimental uncertainty, the pixels in the analyzed ROI can be regarded as statistically independent entropy sources.

Furthermore, to quantify the statistical nature of the extracted fluctuations, we define a noise ratio (NR) metric as the ratio of the measured spatial intensity variance with the total number of photons, as,
\begin{equation}
    \text{NR} \equiv \frac{\Delta^2[(N_{L,n} - N_{L,n+1}) - (N_{R,n} - N_{R,n+1})]}{ \langle N_{L} \rangle + \langle N_{R} \rangle }
    \label{eq:NR},
\end{equation}
where $N_{L,n}$ and $N_{R,n}$ represent the photon counts for the left and right coherent state pulses in frame $n$, respectively. The numerator calculates the spatial variance of the intensity fluctuations obtained via the frame subtraction method, while the denominator represents the mean photon count. 

To account for potential correlations from readout noise, clock-induced charge, or dark current, we performed a noise analysis by spatially binning pixels into super-pixels up to $20 \times 20$. For a purely uncorrelated noise source, the noise ratio remains invariant with respect to the bin size. Any deviation from this invariance directly quantifies residual electronic or technical correlations. The results of the NR as a function of binning are plotted in Fig.~\ref{fig:purity_analysis}(c), which are consistent with Poissonian statistics; the experimental data maintains $\text{NR} \approx 1$ regardless of bin size. This spatial invariance confirms that the extracted fluctuations are governed strictly by intrinsic quantum shot noise, validating the system as a robust and unpredictable entropy source for high-speed quantum random number generation. 

To quantitatively assess the quantum origin of the measured fluctuations, the total spatial noise is expressed as \(
\sigma_{\mathrm{total}}^{2}=\sigma_{\mathrm{shot}}^{2}+\sigma_{\mathrm{bg}}^{2},
\)
where \(\sigma_{\mathrm{bg}}^{2}\) represents the experimentally measured background noise, including detector-related noise and residual classical background contributions. Under the operating conditions employed in this work, the detector operates in a shot-noise-limited regime with a shot-noise to background-noise SNR of approximately \(29\,\mathrm{dB}\), confirming the dominance of quantum shot noise. Moreover, the robustness of the proposed method was further evaluated by analyzing its breakdown regime under increasing classical noise and reduced shot-noise contribution through ROI expansion. 

\begin{figure}[t]
    \centering
    
    \begin{minipage}{0.48\textwidth}
        \setlength{\unitlength}{\linewidth}
        \begin{picture}(1,0.3)
            \put(0,0){\includegraphics[width=\linewidth]{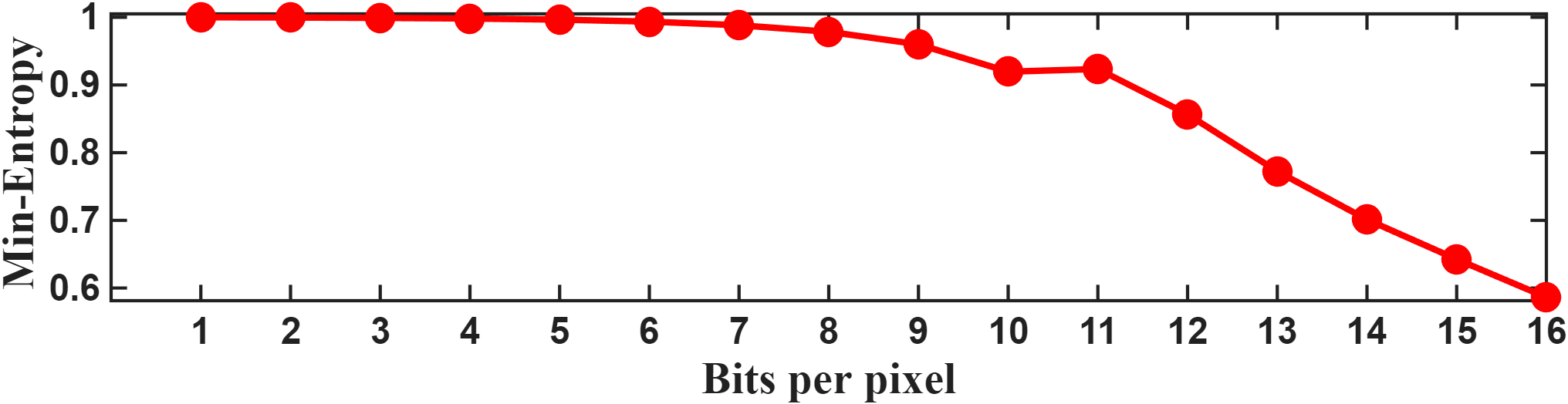}}
            \put(0.11,0.15){\textbf{(a)}} 
        \end{picture}
    \end{minipage}
    \hfill 
    \begin{minipage}{0.48\textwidth}
        \setlength{\unitlength}{\linewidth}
        \begin{picture}(1,0.4)
            \put(0,0){\includegraphics[width=\linewidth]{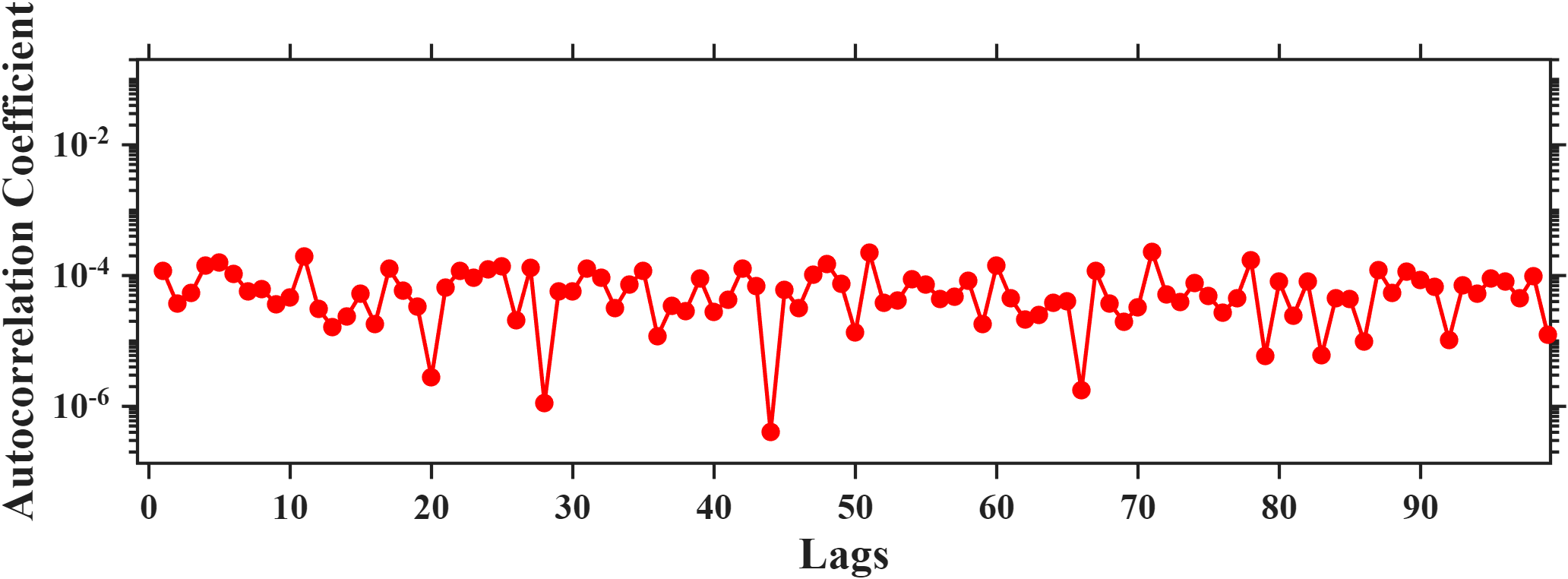}}
            \put(0.11,0.25){\textbf{(b)}}
        \end{picture}
    \end{minipage}
    \caption{(a) Min-entropy $H_{\min}$ as a function of random bit extraction depth. (b) Autocorrelation coefficients as a function of lag, the negligible magnitude of the correlations $(\approx10^{-4}$) validates the statistical independence of the bit extraction process.}
    \label{fig:auto}
\end{figure}

In order to generate a random bitstream, we employed a combinatorial frame-subtraction technique based on a single six-frame kinetic acquisition. Since each acquisition captures two spatially separated beams, all possible frame pairs are processed to generate 30 unique spatial quantum fluctuation images, thereby maximizing the number of random bits extracted from a single kinetic-mode acquisition. The absolute magnitude of the intensity fluctuation images is converted to a 16-bit unsigned integer representation, matching the resolution of the EMCCD's 16-bit analog-to-digital converter. The absolute-value operation is applied solely as a preprocessing step to convert the differential pixel values into non-negative values, and we verified that it does not degrade entropy quality. To determine the optimal random bit extraction depth from the binary sequence, we calculated the min-entropy, defined as $H_{\min}(X) = -\log_2 \left( \max_{x \in \{0, 1\}^n} \Pr[X = x] \right)$, which quantifies the amount of randomness of a distribution X on $\{0,1\}^n$. For a uniform distribution of 0's and 1's, $H_{\min}=1$. We plot the obtained results of min-entropy with number of bits extracted per pixel in Fig.~\ref{fig:auto}(a), and it is clear from the plot that the min-entropy remains near-ideal ($\approx 1$) for about six bit extraction depths and decreases slowly as higher-order bits are included. 

In the present work, we extract the three least significant bits (LSBs) from each pixel to maximize the bit generation rate while ensuring that the entropy originates purely from quantum shot noise.  The extracted bitstreams exhibit a nearly equal distribution of zeros and ones across all three bit positions, enabling the direct generation of a high-entropy and low-bias random sequence without any algorithmic randomness extraction.This approach is based on the principle that when the variance of the quantum-limited distribution significantly exceeds the digitization step size, the lower-order bits become intrinsically equiprobable and statistically uncorrelated\cite{lsb,qelsb}.
\begin{figure}[htbp]
    \centering
    
    \begin{minipage}{0.48\textwidth}
        \raggedright \textbf{(a)} \\
        \vspace{0.2em}
        \includegraphics[width=\linewidth]{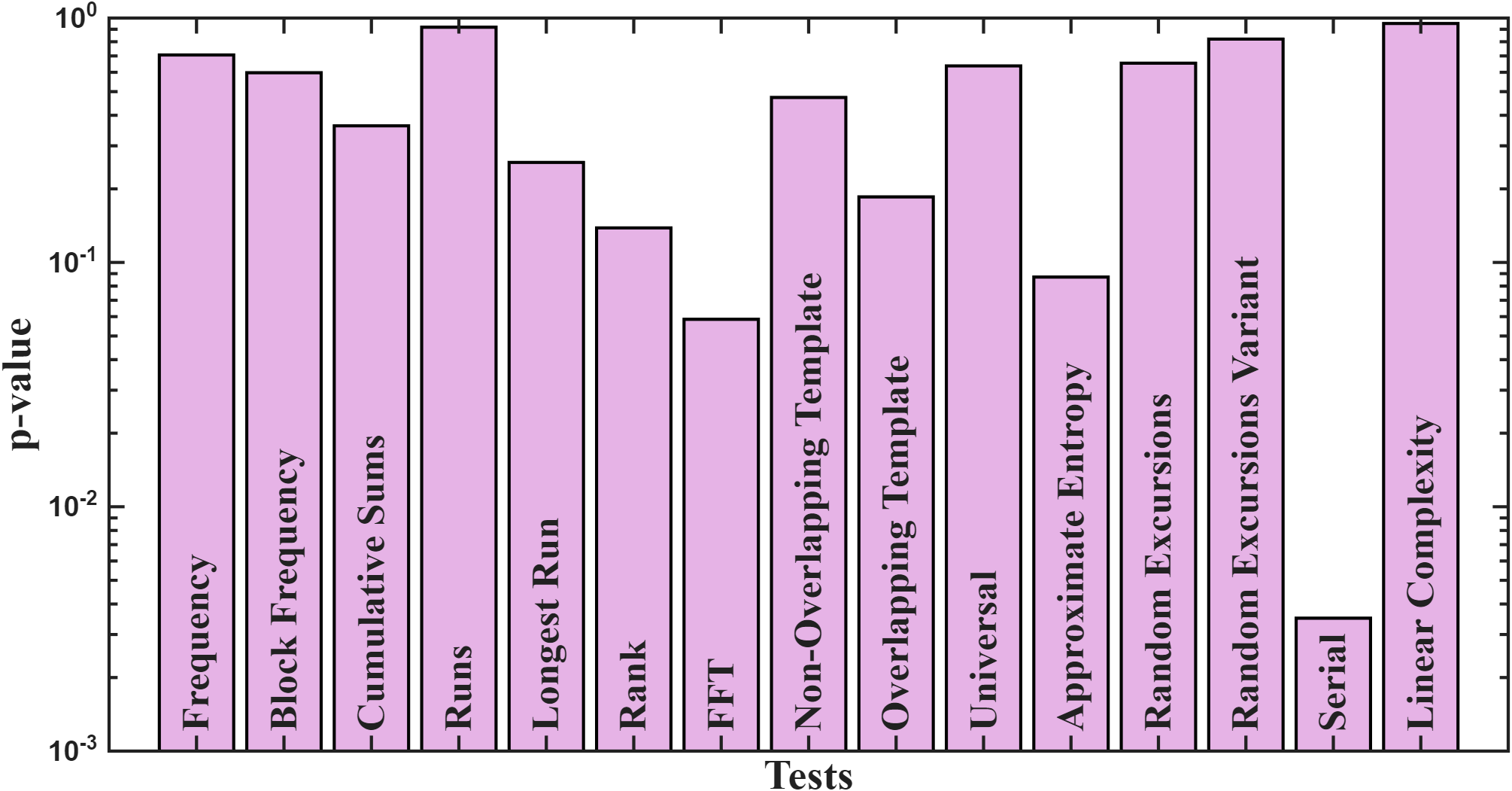}
    \end{minipage}
    \hfill 
    \begin{minipage}{0.48\textwidth}
        \raggedright \textbf{(b)} \\
        \vspace{0.2em}
        \includegraphics[width=\linewidth]{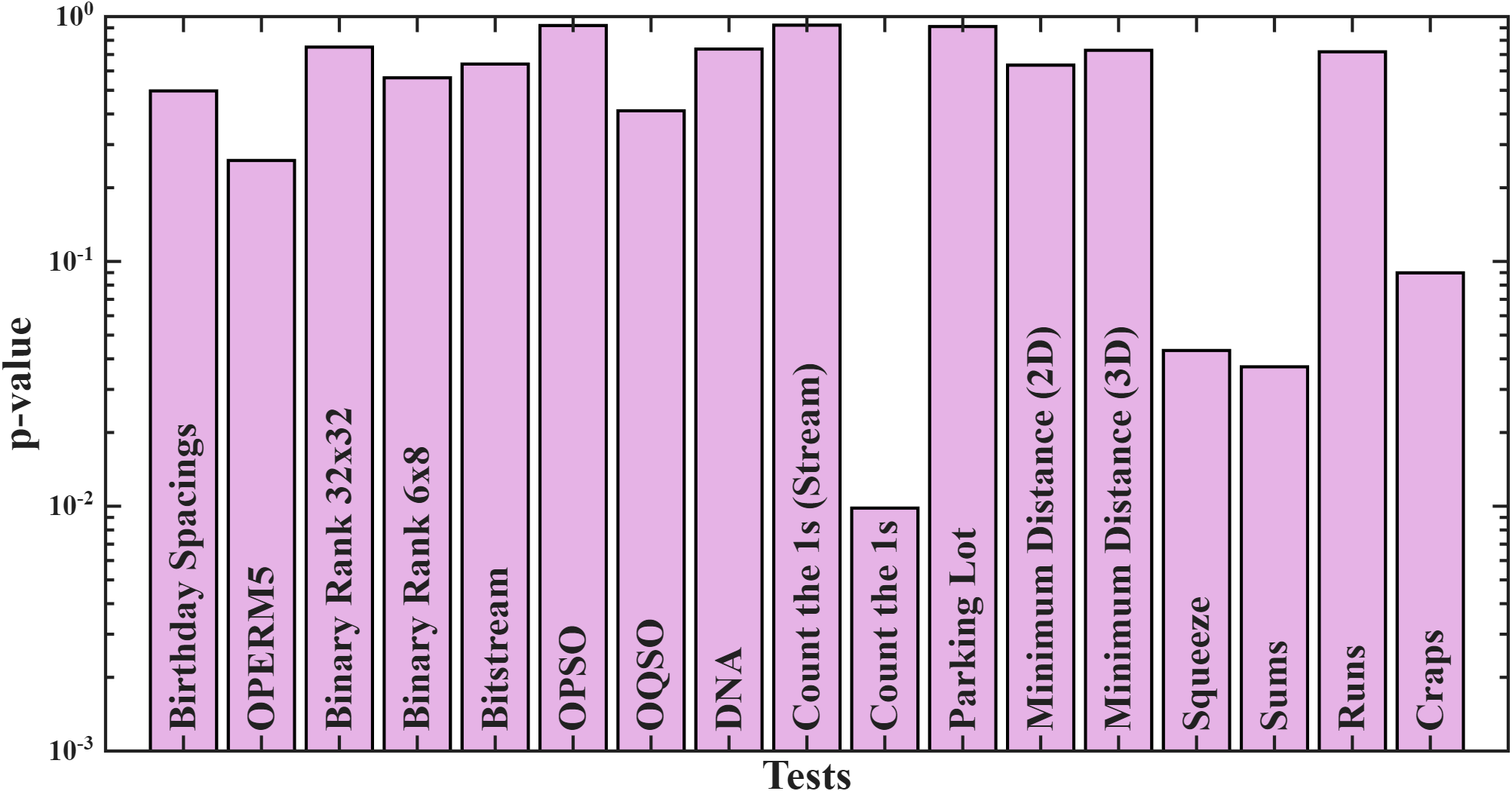}
    \end{minipage}    
    \caption{Statistical validation of the generated bitstream. (a) Performance of the raw bitstream against the NIST statistical suite. The bitstream satisfies the uniformity criterion ($p \geq 0.0001$), confirming the global statistical health of the extraction process. (b) Results of the original Diehard statistical suite.}
    \label{fig:tests}
\end{figure}

Applying this procedure to a $150\times150$ pixel region of each of the 30 combinatorial fluctuation images yielded a total of $2.025\times10^6$ random bits from a single 6-frame acquisition. The generated bitstream exhibits a min-entropy of 0.9966 per bit, while the acquisition time of 342 $\mu$s corresponds to a random bit generation rate of 5.92 Gbps. By scaling the illumination to the full active $170 \times 512$ pixel region of the EMCCD, the intrinsic generation capability expands to a projected rate of approximately 11.45 Gbps.

However, it is critical to distinguish this internal entropy rate from the system's sustained throughput which is governed by the EMCCD's readout architecture rather than the primary entropy source. While the vertical shift registers enable rapid charge transfer rate at 300~ns per row, the total throughput is restricted by the readout electronics. The horizontal amplifier processes pixels serially at a 1~MHz low-noise readout speed, a setting specifically selected to minimize electronic noise and preserve the high-fidelity entropy of the spatial quantum fluctuations. At this frequency, the digitization period is approximately 270~ms per acquisition cycle. Consequently, the continuous generation rate becomes approximately 7.5~Mbps, a value dictated by the sensor's electronic serial-readout speed rather than the intrinsic bandwidth of the quantum entropy source. It is important to note that while the EMCCD supports higher digitization rates up to 20~MHz in EM mode, which could theoretically push the generation rate into the hundreds of Mbps, these higher speeds were avoided in this work to minimize the electronic read noise. Moreover, due to high photon flux of the coherent state pulses present in the system we utilized low noise mode of the EMCCD camera rather than electron multiplying feature. Compared with the conventional QRNG architectures discussed earlier, the proposed EMCCD-based approach offers parallel entropy extraction from thousands of spatial pixels, enabling direct random-bit generation without computational randomness extraction. Its primary limitations are the EMCCD readout speed, the requirement for thermoelectric cooling, and careful detector-noise characterization.

 In order to ensure that the min-entropy remains robust over time, the analysis was scaled across 2,400 acquisitions, totaling $4.86 \times 10^{9}$ bits, leading to min-entropy value to $0.998890$ which remains close to $1$. Furthermore, to verify the non-deterministic nature and statistical uniformity of the generated data, we performed statistical tests over random bits  extracted from 2,400 acquisitions. As a primary validation, the Pearson correlation coefficient for a sample of $10^{8}$ bits was calculated to detect potential short-range correlations in the bitstream. The autocorrelation for the first 100 lags as illustrated in Fig.~\ref{fig:auto}(b) shows auto-correlation coefficients randomly distributed across zero with magnitudes on the order of $10^{-4}$. This confirms the absence of temporal correlations in the sample and validates the stability of the entropy extraction process.

The statistical quality of the generated bitstreams was evaluated using the NIST SP 800-22 statistical test suite, the industry-standard benchmark for cryptographic randomness\cite{nist}. For this validation, the raw, unprocessed bitstream was partitioned into 400 independent sequences, each containing $10^6$ bits. Following standard NIST guidelines, an individual sequence was deemed to pass a given test if its evaluated $p$-value satisfied $p \ge 0.01$. The overall randomness of the bitstream was then verified against two final criteria the proportion of passing sequences had to meet the minimum acceptable rate of 0.9750 for our sample size, and the distribution of the $p$-values was required to exhibit uniformity, defined by a global $p\text{-value} \ge 0.0001$. Tests were conducted using standard parameters, a block length of 128 bits for the block frequency test, 10,000 bits for the overlapping template test, and a bit length of $m=10$ for the approximate entropy test. The results of NIST SP 800-22 statistical test suite is shown in Fig.~\ref{fig:tests}(a), and it can be seen that the global p-values for all 22 statistical test suites fall above 0.0001. 

To ensure a more rigorous evaluation of long-range patterns and subtle distributional defects, we further subjected the bitstream to the Marsaglia Diehard test suite \cite{die}. For tests within the Diehard suite that generate multiple $p$-values across iterations, a Kolmogorov-Smirnov (K-S) test is applied to evaluate their uniformity, yielding a final, single $p$-value. Following established Diehard conventions, a test is considered pass if this resulting $p$-value falls within the stable range of $0.01 \leq p \leq 0.99$. The raw output passed the Diehard tests, as shown in Fig.~\ref{fig:tests}(b). By using this multi layered statistical approach, we ensure that the spatial entropy extracted from our system is robust against both local and global non random patterns.

The present work complements recent source-independent continuous-variable QRNG protocols\cite{di,sicv}. While those approaches prioritize security certification under weaker trust assumptions, our scheme emphasizes experimental simplicity and massively parallel entropy acquisition through spatially resolved detection. Combining these advantages within a source-independent security framework is an interesting direction for future investigation\cite{kumar2021einstein}.

In conclusion, we developed a high-speed quantum random number generator by extracting spatial quantum fluctuations from the intensity profile of the coherent state We first established the fundamental randomness of the source through spatial correlation and shot-noise analyses. Subsequently, the randomness in generated bitstream was rigorously validated through the complete NIST SP 800-22 and Diehard statistical test suites, successfully passing all tests. Currently, the real-time sustained throughput is 7.5 Mbps, while  the system achieves an instantaneous QRNG rate of 5.92 Gbps, which can be scaled up to 11.45 Gbps. The successful extraction of spatial quantum fluctuations validates the present methodology as a robust and highly parallelized framework for high bit rate entropy generation, demonstrating its suitability for rigorous quantum random number generation tasks.
\section*{Acknowledgment}
The authors thank Alberto Marino for valuable discussions on spatial quantum noise extraction and analysis.
\vspace{0.2em}

\end{document}